\documentclass[onecolumn,superscriptaddress,preprintnumbers,pra]{revtex4}
\usepackage{epsfig}
\usepackage{amssymb}
\usepackage{amsmath}
\usepackage{epstopdf}
\usepackage{graphicx,color}  
\usepackage{bm}  
\usepackage{epstopdf}
\usepackage{multirow}

\begin{document}

\title{Scaling limit analysis of Borromean halos}
\author{L.~A Souza}
\affiliation{Instituto Tecnol\'ogico de Aeron\'autica, DCTA, 12228-900, 
S\~ao Jos\'e dos Campos,~Brazil}
\author{F.~F. Bellotti}
\affiliation{Department of Physics and Astronomy, Aarhus University, DK-8000 Aarhus C, Denmark \\
 Instituto de Fomento e Coordena\c c\~ao Industrial, 12228-901, S\~ao Jos\'e dos Campos, 
 SP, Brazil}
\author{T. Frederico}
\affiliation{Instituto Tecnol\'ogico de Aeron\'autica, DCTA, 12228-900, 
S\~ao Jos\'e dos Campos,~Brazil}
\author{M.~T. Yamashita}
\affiliation{Instituto de F\'\i sica Te\'orica, UNESP - Univ Estadual Paulista, C.P. 70532-2, CEP 01156-970, S\~ao Paulo, SP, Brazil}
\author{L. Tomio}
\affiliation{Instituto Tecnol\'ogico de Aeron\'autica, DCTA, 12228-900, S\~ao Jos\'e dos Campos,~Brazil \\ 
Instituto de F\'\i sica Te\'orica, UNESP - Univ Estadual Paulista, C.P. 70532-2, CEP 01156-970, S\~ao Paulo, SP, Brazil}
\date{\today }

\begin{abstract}
The analysis of the core recoil momentum distribution of 
neutron-rich isotopes of light exotic nuclei is performed within a model 
of the halo nuclei described by a core and two neutrons dominated by 
the $s-$wave channel. We adopt the renormalized  three-body model with a 
zero-range force, that accounts for the universal 
Efimov physics. This model is applicable to nuclei with large two-neutron halos 
compared to the core size, and a neutron-core scattering length larger 
than the interaction range.  The halo wave function in 
momentum space is obtained by using as inputs the two-neutron separation 
energy and the energies of the singlet neutron-neutron  and neutron-core virtual states. 
Within our model, we obtain  the momentum probability densities for the
Borromean exotic nuclei Lithium-11 ($^{11}$Li), Berylium-14 ($^{14}$Be) and 
Carbon-22 ($^{22}$C). A fair reproduction 
of the experimental data was obtained in the case of  the core recoil 
momentum distribution of  $^{11}$Li and $^{14}$Be, without free parameters.
By extending the model to $^{22}$C, the combined analysis of the core momentum 
distribution and matter radius suggest (i) a $^{21}$C virtual state well below 1 MeV; 
(ii) an overestimation of the extracted matter $^{22}$C radius; and 
(iii) a two-neutron separation energy between 100 and 400 keV.
\end{abstract}

\maketitle

\section{Introduction}\label{sec:introduction}

The large reaction cross section of Carbon-22 ($^{22}$C) observed by Tanaka and collaborators 
\cite{TanPRL10}, in the collision of $^{22}$C on a liquid hydrogen target at 40A MeV compared to 
the neighbour carbon nuclides, $^{19}$C and $^{20}$C, triggered a theoretical 
discussion \cite{YamPLB11,ForPRC12,ErsPRC12,AchPLB13,AchFB15} about the
possible constraint that the associated extracted matter root-mean-square (rms) radius 
$\sqrt{\langle r_m^2\rangle}$ can have on the two-neutron separation energy, $S_{2n}$. 
From the experimental point of view the two-neutron separation energy of $^{22}$C
is not well known. A recent atomic mass evaluation (Ame2012), reported in 
Ref.~\cite{AudiCPC2012-II}, gives $S_{2n}=$ 110 (60) keV for  $^{22}$C. And, from a
mass measurement~\cite{GauPRL12}, $S_{2n}=$ -0.14 (46) keV.
There is also an indirect evidence that $^{22}$C could be bound by less 
than 70 keV~\cite{MosNPA13}.

Such reaction cross-section, analysed with a finite-range Glauber 
calculation under an optical-limit approximation, suggested that  
$\sqrt{\langle r_m^2\rangle}\,=\,5.4\,\pm\,0.9$~fm~\cite{TanPRL10}. 
Their analysis showed that the two-valence neutrons prefer to occupy the  
$s_{1/2}$ orbital. Considering the observed $^{22}$C matter 
radius together with  the $^{20}$C one, it was given in Ref.~\cite{YamPLB11}
the following estimate for the radius of the neutron halo orbit:
\begin{equation}
\sqrt{\langle r^2_n \rangle}=\sqrt{\frac{22}{2}}\sqrt{\langle r_m^2[^{22}C]\rangle-
\frac{20}{22}\langle r_m^2[^{20}C]\rangle} \,= \,15\,\pm\,4 \, \text{fm}
,\end{equation}
where $\sqrt{\langle r_m^2[^{20}C]\rangle}\,=\,2.98(5)$ fm\cite{OzaNPA01}.
The $^{22}$C rms matter radius is used as a constrain to the poorly known 
experimental value of the two-neutron separation, in a renormalized zero-range
three-body model~\cite{YamPLB11}, in a shell model approach~\cite{ForPRC12}, 
in a three-body cluster model~\cite{ErsPRC12}, and  within 
an effective field theory  with contact interactions~\cite{AchPLB13}. All these
 approaches agree to a value around or below 100 keV. However, independent 
 information are relevant to cross check such estimations. 
 Indeed, in Ref.~\cite{ErsPRC12},  a strong correlation between the soft dipole mode in 
 the Coulomb fragmentation  process of $^{22}$C and $S_{2n}$ was also shown.

Comparison between the experimental data for the core recoil momentum distributions 
of $^{22}$C~\cite{KobPRC12} and Lithium-11 ($^{11}$Li)~\cite{TanJPG96} indicates  
similar sizes for their two-neutron halos, which suggest that the matter radius of the carbon isotope could be 
overestimated (see Ref.~\cite{RiiPST13}, for further discussion). 
This motivates us to study the properties of low-momentum distributions in two-neutron 
halo light exotic nucleus, which are dominated by $s-$wave short-range two-body interactions 
in the neutron-neutron ($n-n$) and neutron-core ($n-c$) subsystems.

In the present work, 
the halo properties are analysed by using correlation between observables or 
scaling functions, involving the width of the core recoil distribution, 
$S_{2n}$, the energies of the $s-$wave virtual states of the singlet spin 
$n-n$ system and $^{21}$C. Furthermore, correlations between the 
matter radius and the quantities mentioned before are also used in our combined 
analysis of the existing data for the width,  matter radius and the virtual state of 
$^{21}$C. Our interest is to constraint the parameters associated with the halo 
structure and two-neutron separation energy based on these data fitted to 
three-body model calculations, by adding to the work presented in Ref.~\cite{SouarXiv15} 
another universal correlations of the width of the core momentum distribution and 
$^{22}$C matter radius.

The model calculations for the core recoil momentum distribution are 
performed within a zero-range three-body model applied to a neutron-neutron-core 
($n-n-c$) system, that fully accounts for the so-called 
Efimov physics \cite{efimov1,efimov2,efimov3}.
The detailed formulation of the momentum 
distributions was given in~\cite{YamPRA13} and the inputs of the renormalized 
zero-range model are the scattering lengths, and one three-body scale, which is given by 
the two-neutron separation energy. Therefore, three low-energy 
observables are enough to determine another low-energy quantity through an 
appropriate universal scaling function computed within the zero-range model. 
Following that, we established the above mentioned correlations between two $s-$wave 
halo observable, like, e.g., the width of the momentum distribution with $S_{2n}$,
matter radius with $S_{2n}$, and the width and matter radius. In particular, this last
one  was useful to constrain the virtual state energy of $^{21}$C, as will be 
discussed in detail.

In our work, the data for the core recoil momentum distribution extracted from the halo 
fragmentation reaction, is interpreted on the basis of the Serber model \cite{SerPR47},
where the peak around zero momentum is proportional to the momentum distribution. 
We test the reliability of our analysis with $^{11}$Li (see e.g. \cite{TanJPG96}) and 
$^{14}$Be, as the low energy parameters used as input
to our calculations are reasonably known. The recent experimental 
results for $^{20}$C and $^{22}$C \cite{KobPRC12} allow us, in principle, to constraint 
$S_{2n}$ and the matter radius of $^{22}$C. In addition, we consider the constrain to the 
virtual state of $^{21}$C by also computing the matter radius, which has a given value in
Ref.~\cite{TanPRL10}. An issue that is beyond the present work, is the consideration of 
the range corrections to the scaling functions, either by calculations with 
finite range potentials  or along the lines suggested in Ref.~\cite{HadiPRA13}, 
as well as in a series of recent works \cite{GatPRA12,KiePRA13,KiePRA14,JiPRA15},
which we leave for a future investigation.

The paper is organized as follow. In section 2, the experimental
$^{22}$C core recoil momentum distribution
is revised and the width associated with the two-neutron halo is singled out.
 In section 3, the scaling relations for the $n-n-c$ system close to the unitary limit is presented.
In section 4, the scaling analysis for the $^{22}$C halo is performed in light of the
experimental result for the momentum width. Finally, in section 5, we have our summary.

\section{ Core recoil momentum distribution in $^{22}$C: phenomenological analysis  }\label{sec:crmd}

The information on the core recoil momentum distribution in two-neutron halo nuclei is 
assessed through the removal cross-section from the collision of these radioactive exotic 
nuclei with nuclear targets at few hundred MeV/nucleon (see e.g. \cite{TanPPNP12}). For 
neutrons flying out from  $s-$states, these removal cross-sections shows 
a characteristic narrow peak at zero core recoil momentum, and a width that features the 
weak binding of the neutrons to the core. In the example of $^{11}$Li \cite{TanJPG96}, it 
is visible the contribution of distributions with a narrow and wide widths. The narrow 
momentum distribution is associated with the two neutrons in the halo state, and the
wide one to the neutrons in the inner orbits in $^{11}$Li and inside the $^9$Li core. 
Adopting such interpretation, the final core momentum distribution is an incoherent 
sum of two distributions one narrow and another wide. Of course this picture is based on 
a large separation between the halo neutron orbit and the inner ones, implying in a small 
overlap among them, and suppression of the coherent terms in the removal cross-section. By
scaling arguments, the width of the narrow peak in $^{11}$Li should be of the order of 
$\sqrt{m_n\,S_{2n}}=18$ MeV/c, where $m_n$ is the nucleon mass.
 This is  quite close to the experimental fit, $21(3)$MeV/c, given in 
 Ref.~\cite{TanJPG96}. For such estimate we use the
two-neutron separation energy of $S_{2n}= 369.15(65)$ keV~\cite{SmiPRL08}.

\begin{figure}[h]
\vspace{1.2cm}
\begin{center}
\includegraphics[width=7.7cm]{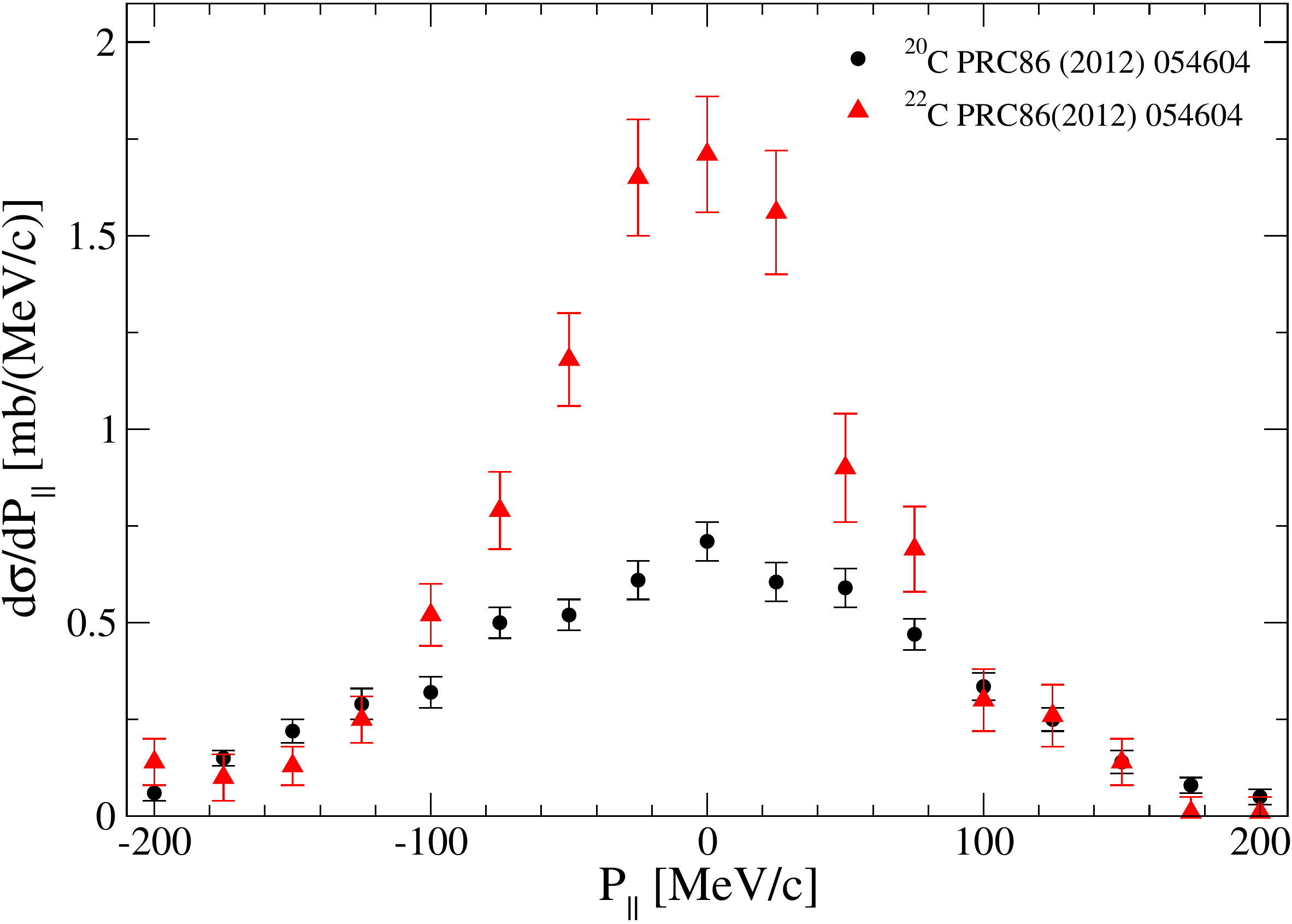}
\includegraphics[width=7.7cm]{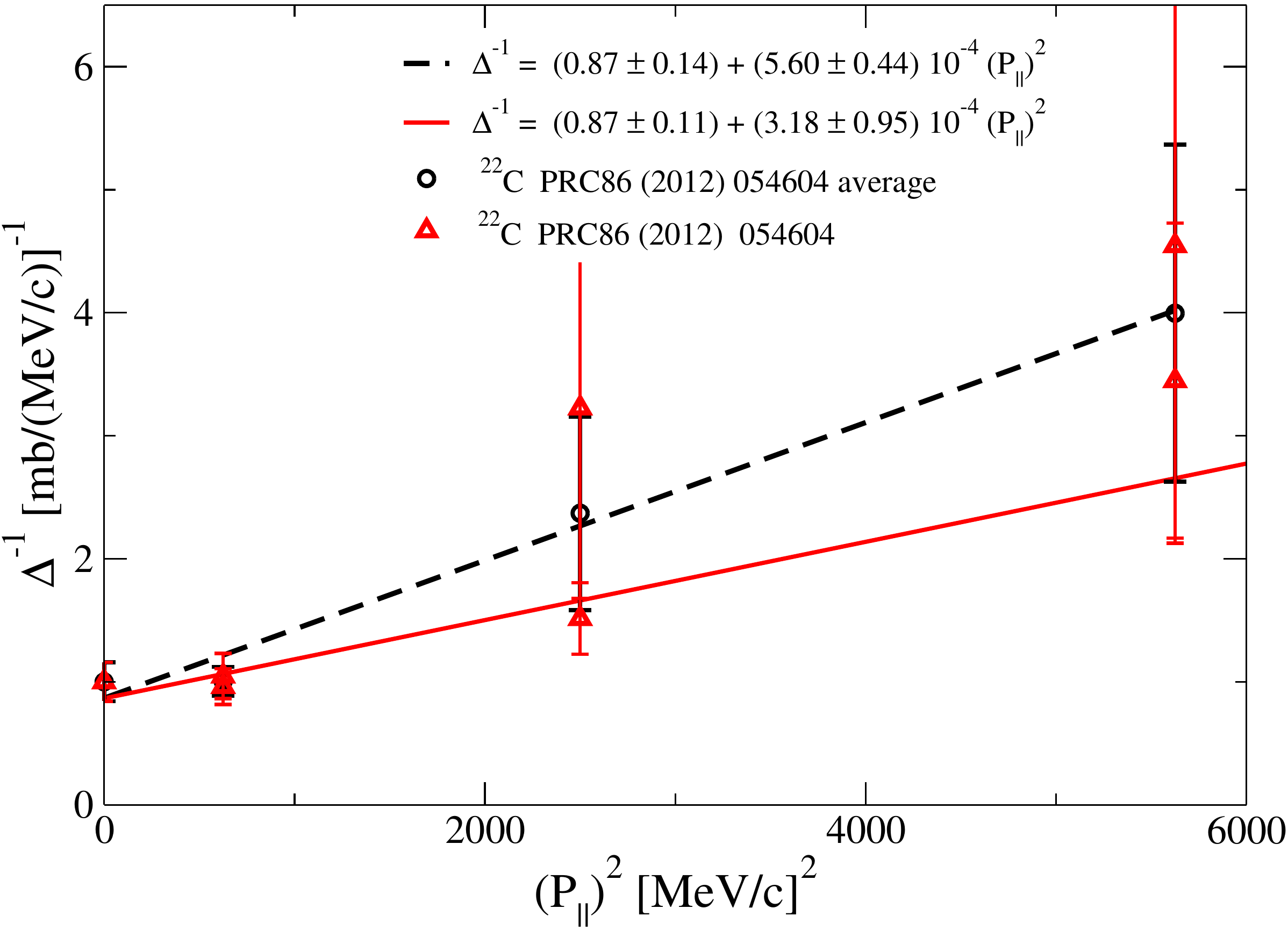}
\end{center}
\caption{(color online). {\it Left-frame:} Experimental data for two-neutron removal cross-section from $^{22}$C
(red-triangles) and from $^{20}$C (black-circles), obtained in a collision with a carbon target at 240 MeV/A, given as a function of the rest frame 
core recoil momentum \cite{KobPRC12}.
 {\it Right-frame:}
The inverse of the difference between the removal cross-sections of $^{22}$C and $^{20}$C,
 given by Eqs.~(\ref{d0}) (black-dashed line) and (\ref{d1}) (solid-red line)
 are shown as functions of the squared recoil core momentum. Circles (triangles)
 represent the averaged difference  for $P_{||}>0$ and $P_{||}<0$, fitted by the dashed (solid) straight line.}
\label{fig:c22-1}
\end{figure}

The above description applied to  $^{22}$C as formed by two neutrons in a weakly-bound 
halo with a core of $^{20}$C, one expect that the core recoil momentum 
distribution presents itself as a narrow peak on the top of a wide one, once the halo 
neutrons occupy an $s$ state. The narrow and wide peaks should be associated with well 
separated length or momentum scales, in order to characterize a true two-neutron halo 
state. We can resort to the recent experimental information on two neutron 
removal cross-section \cite{KobPRC12} of $^{22}$C and $^{20}$C colliding with a carbon 
target at 240 MeV/A. The experimental resolution for both cases are very close. 
This allows to make the comparison between the experimental removal 
cross-sections as a function of the core momentum as shown in the left panel of 
Fig.~\ref{fig:c22-1}. Apart small kinematical corrections, it is quite clear
that the wide distribution corresponds to neutrons removed from the core of 
$^{22}$C, and the narrow peak is clearly characterized. 

To express such separation among the length/momentum 
scales of the halo state and the inner core orbits in $^{22}$C, 
we introduce a phenomenological quantity, which gives the 
difference the two-neutron removal cross-sections in $^{22}$C and $^{20}$C
\begin{equation}
\Delta \left(P^2_{||}\right)=\frac{d\sigma}{dP_{||}}[{^{22}C}]-\frac{d\sigma}{dP_{||}}[{^{20}C}] \ ,
\end{equation}
which should be associated with the two-neutron halo property of  $^{22}$C, 
as the reaction rate associated with the removal of neutrons from the inner orbits is expected to be the same 
as in $^{20}$C. 
The core recoil momentum distribution is parametrized as a normal distribution 
\begin{equation}
\Delta \left(p^2\right)=   N \, \exp\left({\displaystyle -\frac{p^2}{2\, {\rm s}^2}}\right) \ ,
\end{equation}
where the variance is given by ${\rm s}^2$, with the corresponding standard deviation
${\rm s}=\sqrt{{\rm s}^2}$ associated with the Full Width at Half Maximum (FWHM) distribution, 
$w\equiv \sqrt{8 \ln 2}\;{\rm s}$. $N$ is a normalization from the data.
 In the right frame of Fig.~\ref{fig:c22-1},
we plot the data for $\Delta^{-1}$ as  a function of $P^2_{||}$. 
We also present the averaged data for positive and negative $P_{||}$, shown by empty circles. A straight line fit is suggested by data.
For $p$ given in units of MeV/c, with $p\equiv ({\bar p})$ MeV/c, considering the averaged experimental 
values, we obtain
 \begin{equation} \label{d0}
\Delta^{-1} (p^2)\,
= \left\{(0.87 \pm 0.14) + (5.60 \pm 0.44)\, 10^{-4}\,{\bar p}^2 \right\} \left[\text{mb/(MeV/c)}\right]^{-1} .
\end{equation}
The result for the FWHM
is $w=79\, \pm \, 22$ MeV/c, with the associated ${\rm s}_{av}=  34\, \pm \, 9$ MeV/c. Taking into account 
the experimental resolution of ${\rm s}_{exp}= 27$ MeV/c \cite{KobPRC12}, we obtain 
${\rm s}_{av}=  20\, \pm \, 15$ MeV/c. 
The fitting of the difference  without the average is given by
\begin{equation}\label{d1}
\Delta^{-1}(p^2)\,= \left\{(0.87 \pm 0.11) + (3.18 \pm 0.95)\, 10^{-4}\,{\bar p}^2 \right\}
\left[\text{mb/(MeV/c)}\right]^{-1} \ ,
\end{equation}
such that for the FWHM we have $w=104\, \pm \, 39$ MeV/c, with the corresponding 
${\rm s}=  44\, \pm \, 16$ MeV/c. After considering the experimental resolution this value is
reduced to ${\rm s}=  34\, \pm \, 21$ MeV/c. The FWHM quoted in \cite{KobPRC12} for 
the narrow peak of $^{22}$C is 73 MeV/c, with ${\rm s}=\,31$ MeV/c, compatible with 
our phenomenological analysis.
 
Naively, one could estimate that the narrow peak leads to $S_{2n}\sim {\rm s}^2/m_n$. In this
case, for ${\rm s}=$ (20, 34, 31) MeV/c the corresponding values of $S_{2n}$ are given by 
(0.4, 1.2, 1) MeV. These values appear to be somewhat large  according to
the analysis performed in \cite{YamPLB11,ForPRC12,ErsPRC12,AchPLB13}, where the suggested 
upper limit for $S_{2n}$ is  $\sim$ 0.1 MeV, to achieve consistency with 
the extracted rms matter radius of 5.4 $\pm$ 9 fm \cite{TanPRL10}. 
In order to refine our estimate of $S_{2n}$, we 
will rely on three-body calculations in the zero-range limit, where the universal 
scaling laws are  defined and computed, and in addition the errors 
from our fitting procedure have to be considered.
   
\section{The neutron-neutron-core system close to the unitary limit}\label{sec:nnc}

Our goal here is to present an analysis of the core recoil momentum distributions 
of light two-neutron halo-nuclei based on  scaling laws computed for 
large scattering lengths, close to the unitary limit and dominated by the universal 
Efimov physics \cite{efimov1,efimov2,efimov3}.  
We adopt a renormalized three-body zero-range model, 
with the halo nucleus described as two neutrons 
with an inert core ($n-n-c$), which 
is appropriate for the analysis of such weakly-bound systems, where the 
$n-n$ and the $n-c$ interactions are dominated by $s-$wave states 
(see e.g. \cite{FreFBS11,FrePPNP12,ZinJPG13}). 
Examples of two-neutron halo nuclei, which
fits on these assumptions, are $^{11}$Li, $^{14}$Be and the carbon systems $^{20}$C 
and $^{22}$C. We study the width of the core recoil momentum distribution and make use of 
the existing data for $^{11}$Li~\cite{TanJPG96} and $^{14}$Be~\cite{ZahPRC93} to validate 
the model, and then analyse the case of
$^{22}$C~\cite{KobPRC12}. The momentum distribution is extracted 
from the halo fragmentation on nuclear targets (see e.g. Ref.~\cite{TanPPNP12}).
The present analysis detail our recent study \cite{SouarXiv15} and emphasizes 
the $^{22}$C case. 

The core recoil momentum distribution  in  the $n-n-c$ system is given by:
\begin{eqnarray}
\label{nqaqb}
n(q_c)=\int d^3p_c |\langle\vec{q}_c\vec{p}_c|\Psi\rangle|^2,
\end{eqnarray}
where the relative momentum of the core to the $n-n$ subsystem is $\vec q_c$ and 
the relative momentum between the neutrons is $\vec p_c$, which is integrated above. 
Detailed expressions are given in~\cite{YamPRA13}, within the renormalized three-body 
zero-range model. It requires as inputs the scattering lengths for the 
$n-c$ and $n-n$ systems, besides one three-body ($n-n-c$ ) scale, which will be 
chosen as $S_{2n}$. In the unitary limit, besides the core mass $A$, the 
only scale that remains is the three-body binging energy. The geometrically separated 
tower of Efimov states is completely determined by such energy, as a necessary 
requirement from the relation between the Efimov effect and the Thomas collapse \cite{AdhPRA88}. 
In what follows, instead of the scattering lengths, we are going to refer to the 
corresponding   two-body virtual state energies (as the present study is limited to
Borromean $n-n-c$ systems), which are taken by their absolute (positive) quantites
 $E_{nc}$ and $E_{nn}$. The $n-n$ virtual state energy in $s-$wave is known 
 and in general will be fixed to 143 keV.

The momentum space $s$-wave  three-body wave function ($|\Psi\rangle$), for the 
renormalized zero-range three-body model applied to the two-neutron halo nuclei, 
can be written in terms of two spectator functions, 
$\chi_{nn}(q_c)$ and $\chi_{nc}(|\vec{p}_c\pm\frac{\vec{q}_c}{2}|)$ \cite{FrePPNP12}, 
where the bound-state three-body energy is replaced by the two-neutron separation
energy $S_{2n}$, such that
\begin{eqnarray}\label{eqsw1}
\langle\vec{q}_c\vec{p}_c|\Psi\rangle=
\frac{\chi_{nn}(q_c)+\chi_{nc}(|\vec{p}_c-\frac{\vec{q}_c}{2}|)
+\chi_{nc}(|\vec{p}_c+\frac{\vec{q}_c}{2}|)}{S_{2n}+H_0}.
\end{eqnarray}
In the above, the corresponding free Hamiltonian is given by 
\begin{eqnarray}\label{h0}
H_0\equiv\frac{p_c^2}{2m_{nn}}+\frac{q_c^2}{2m_{nnc}},
\end{eqnarray}
where $m_{nn}$ and $m_{nnc}$ are the reduced masses, given by 
$m_{nn}={m_n}/{2}$ and $m_{nnc}=2{A}m_n/(A+2)$, with ${A}\equiv {m_c}/{m_n}$.
The core is assumed structureless with the two neutrons in a singlet spin state to ensure the 
antisymmetrization of the halo wave function. 
We observe that, as large is the halo neutron orbit, better is the assumption of a structureless 
core from the point of view  of the full antisymmetrization of the nuclear wave function.

The spectator functions in the renormalized three-body model 
are the solution of a coupled set of homogeneous integral equations, which 
correspond to the Faddeev equations for the zero-range interaction 
subtracted at a given scale. 
Besides the $n-n$ and $n-c$ energies, within the zero-range model we need to provide
a regularization scale, which is given by an energy parameter $-\mu^2$.
This regularization is done by fixing $S_{2n}$. 
The value of $\mu$ can be moved towards infinity, generating a tower of Thomas-Efimov 
excited states.
The log-periodic behaviour of the spectator functions appear in the ultraviolet region, 
not affecting the low momentum part of the wave function 
that indeed runs to a limit cycle. This behaviour is explored in 
detail in \cite{YamPRA13}. It was also found an universal ratio 
between the magnitude of the log-periodic spectator functions in the limit of 
$\mu\to\infty$, which is a function of only the mass ratio $A$. We noticed that
in practice the first cycle is enough to get the properties of the wave function 
in the infrared momentum region, which is acceptable for our purposes of computing the
width of the momentum distribution. This happens because in the actual cases of for 
example, $^{11}$Li and $^{22}$C, one finds that $\mu^2 \gg S_{2n}$. Therefore, 
our calculation of the scaling law for the width of the core recoil momentum 
distribution  is close to the universal limit cycle already in the first cycle. 
We have checked in \cite{SouarXiv15} that correlation between the 
width and $S_{2n}$ of the ground, first 
and second excited states are quite close and in practice the limit cycle for this 
scaling law is achieved with the ground state calculation, of
the renormalized zero-range three-body model~(see e.g. \cite{FrePPNP12}), 
namely given by the subtracted 
Skorniakov and Ter-Martirosian equations for mass imbalanced systems~\cite{SKT}.

The utility of a scaling function in few-body physics was recognized long ago by 
Phillips~\cite{PhiNPA68} in the analysis of the doublet neutron-deuteron scattering 
length through the correlation with the triton binding energy, which he found quite model 
independent. In the 80's Efimov showed that the origin of the model independence of the
``Phillips line" is the universal long range potential, which emerges close to the unitary 
limit (infinite two-body scattering length). Our approach just follows that idea,
searching for the correlation between any other two three-body low-energy $s-$wave 
observable, by using the zero-range model to compute the associated scaling law.  One 
can use a scaling function to estimate for example the value of $S_{2n}$ or ${\rm s}$, 
knowing one of them and eventually use it to constrain some other poorly known 
low-energy observable, such as a subsystem scattering length.

The scaling function for the width of the core recoil momentum distribution ${\rm s}$,
gives a correlation between ${\rm s}$ and the energies $S_{2n}$, $E_{nc}$ and $E_{nn}$,
which can be written in a general form as
\begin{equation}
{\rm s}\,=\, \sqrt{S_{2n}\,m_n}\,{\cal S}_c\left(\pm\sqrt{\frac{E_{nn}}{S_{2n}}},\,
\pm\sqrt{\frac{E_{nc}}{S_{2n}}}; {A}\right) \, 
,\label{sigmac}
\end{equation}
where the $+$ and $-$ signs refer to the bound and virtual subsystem energies, respectively. 
In the present specific Borromean three-body case, we have only virtual two-body subsystems,
such that both signs are negative in Eq.~(\ref{sigmac}), with $E_{nn}$ fixed to the $n-n$ 
virtual state.
For the momentum distribution width, the scaling function (\ref{sigmac}) 
is the limit cycle of the correlation 
function associated with ${\rm s}$ as a function of 
$E_{nn}$, $E_{nc}$ and $S_{2n}$, when the three-body ultraviolet (UV) cut-off 
is driven to infinite in the three-body integral equations, or equally the
subsystem energies are let to zero with a fixed UV cut-off. 
Similar procedure is performed within a renormalized zero-range three-body model, 
in the subtracted integral equations, where the subtraction energy is fixed and the
two-body scattering lengths are driven towards  infinite. 
In practice, both procedures provides very close results, as shown 
in Ref.~\cite{YamPRA02}. In the exact Efimov limit ($E_{nn}=E_{nc}=0$), 
the width is a universal function of the  mass number $A$, 
\begin{equation}\label{s0}
{\rm s}\,=\,\sqrt{S_{2n}\,m_n}\, {\cal S}_c\left(0,\,0, { A}\right)\approx
1.154\,\sqrt{S_{2n}\,m_n}\,\frac{\sqrt{A}/\pi\,+\,A}{1\,+\,\sqrt{A}/\pi\,+\,A } \ .
\end{equation}
The scaling function is associated to the limit cycle of the correlation between ${\rm s}$ 
and $S_{2n}$, computed in principle for an Efimov state with large excitation. 
The constant $\pi$ in the above formula was found by the fitting procedure to 
the numerically calculated scaling function given in \cite{SouarXiv15}.
\begin{table}[h]
\begin{center}
\begin{tabular}{cccc}
\hline\hline \\
 Borromean Nuclei & 
$S_{2n}$ [keV]  \cite{AudiCPC2012-II}  & ${\rm s}_{th}$ [MeV/c]  &  ${\rm s}_{exp}$ [MeV/c]   \vspace{0.2cm}\\
  \hline\hline\\
 $^{11}$Li & 369.28$\pm$0.64  &  19.5  &  21$\pm$3 \cite{TanJPG96} \vspace{0.1cm}\\
 $^{14}$Be & 1265.89$\pm$132.26 & 37$\pm$2 &  39.3$\pm$1.1 \cite{ZahPRC93} \vspace{0.2cm}\\
 $^{22}$C & 110$\pm$60 & 11$\pm$3 & 31 \cite{KobPRC12}; $20\, \pm \, 15$ (\ref{d0}); $34\,\pm\,21$ (\ref{d1}) 
 \vspace{0.1cm}\\
\hline\hline
\end{tabular}
\end{center}
\caption{Theoretical values of ${\rm s}_{th}$, obtained from the scaling function (\ref{s0}), are compared with
the corresponding experimental values, ${\rm s}_{exp}$, which were extracted from the core recoil momentum 
distribution of the Borromean halo nuclei $^{11}$Li, $^{14}$Be and $^{22}$C.
In the case of $^{22}$C, we consider the averaged expression (\ref{d0}), as well as (\ref{d1}) without 
averaging.}
\label{tab:tab1}
\end{table}

In Table \ref{tab:tab1} we compare the results of Eq. (\ref{s0}) for ${\rm s}$ in the case
of the Borromean halos $^{11}$Li, $^{14}$Be and $^{22}$C. 
The comparison is illustrative as we are considering only the unitary limit, and therefore
disregarding the small virtual state energies of the subsystems. The dependence of 
${\rm s}$ on the subsystems energies are weak, and small corrections are present, as we 
have verified in \cite{SouarXiv15}. In the case of $^{22}$C the calculation is poor 
compared to the experimental value of ${\rm s}$ from \cite{KobPRC12}, 
while  our phenomenological analysis with values of ${\rm s}$ obtained from the fits 
(\ref{d0}) and (\ref{d1})  are able to
overlap with the theoretical estimation due to the large uncertainties. 
We will present more results  in the next section and  return to this discussion,
despite the small effects on ${\rm s}$ from $E_{nn}$ and $E_{nc}$ for a given two
 neutron separation energy. We point out that, the model is validated by 
 the comparison with 
 the experimental values in the cases of $^{11}$Li and $^{14}$Be. 

\begin{figure}[h]
\vspace{1.2cm}
\begin{center}
\includegraphics[width=7.7 cm]{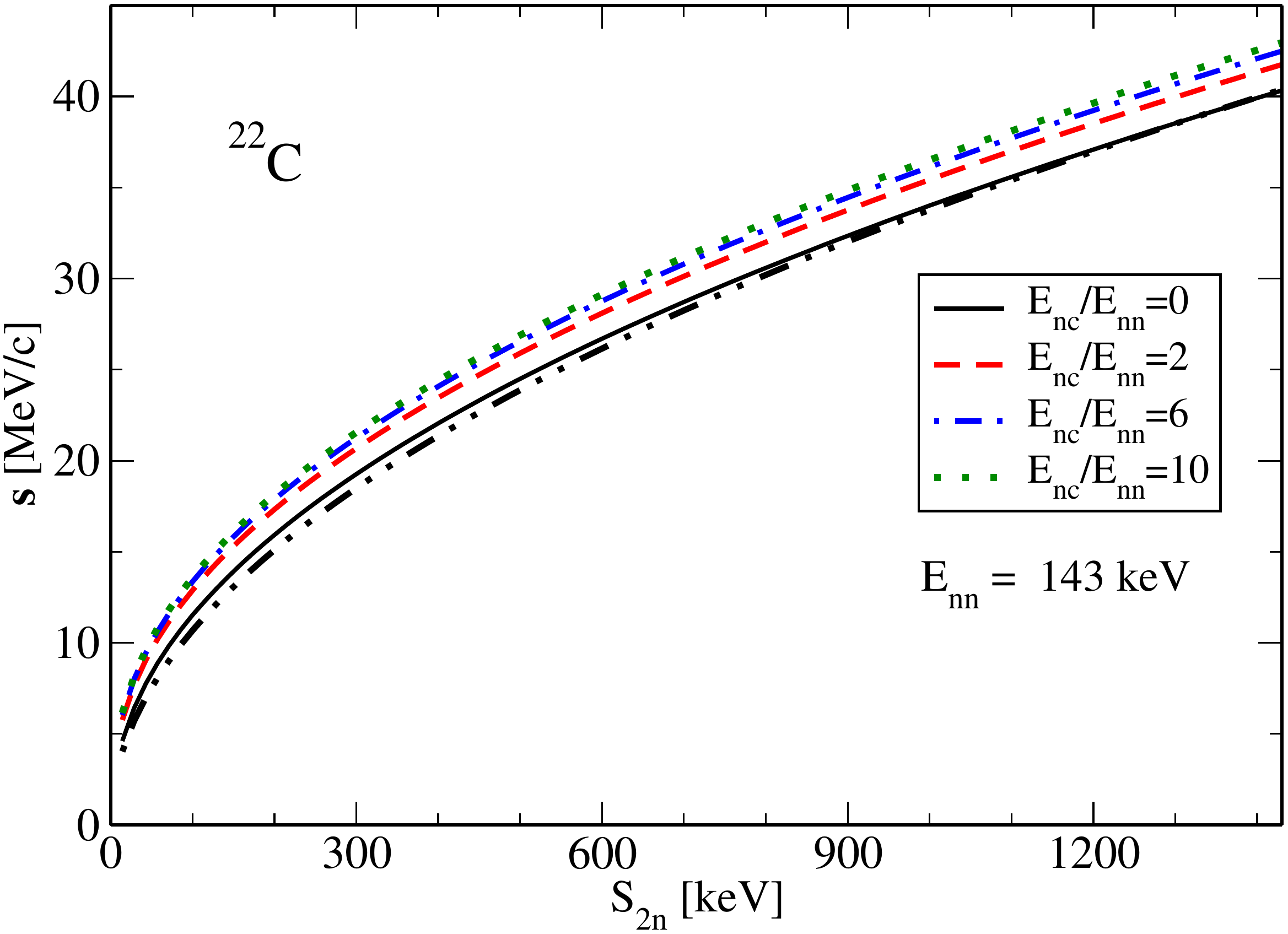}
\includegraphics[width=7.7 cm]{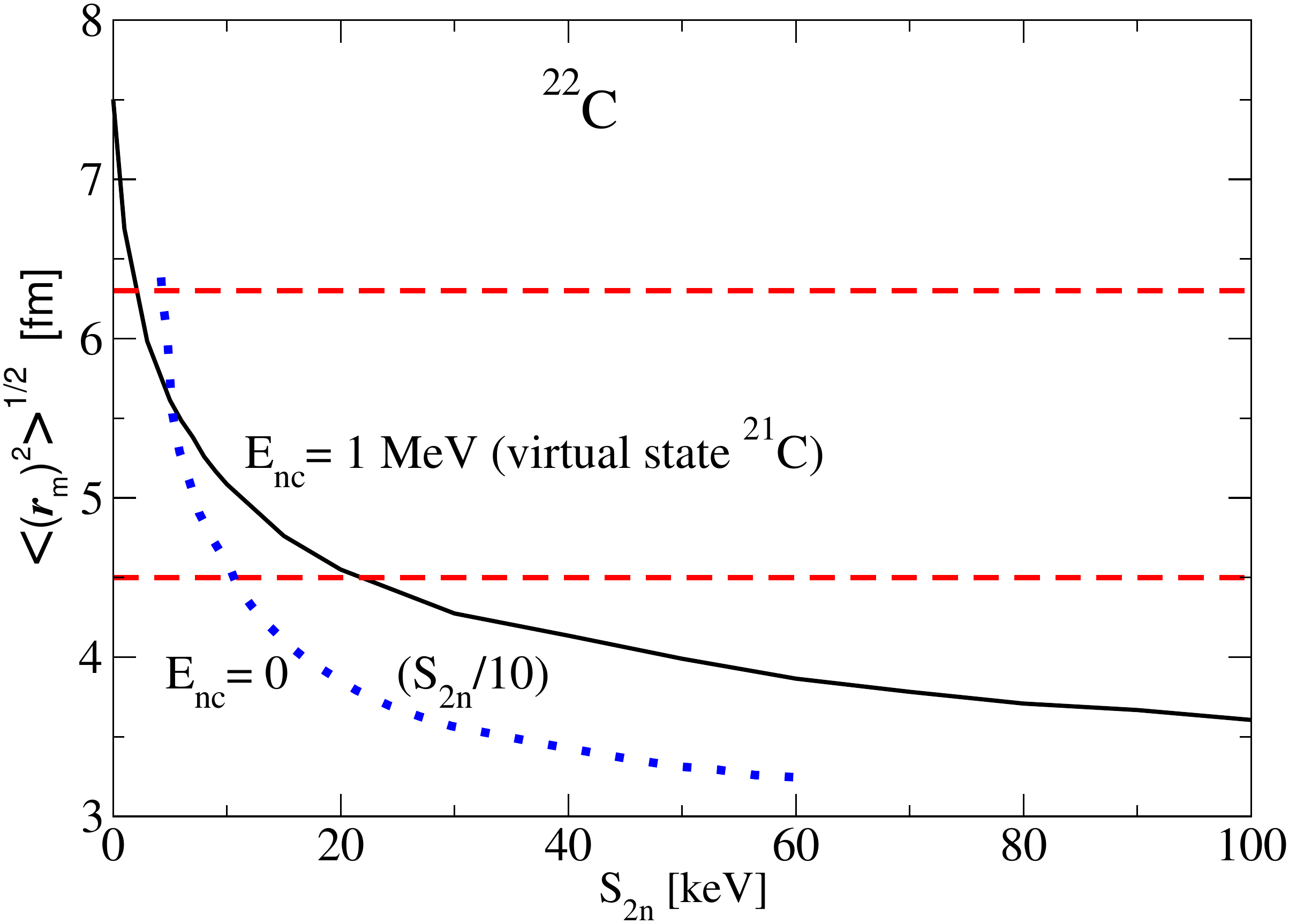}
\end{center}
\caption{(color online). {\it Left-frame:} Scaling plot for the width ${\rm s}$ as a function of $S_{2n}$, for 
$^{22}$C with different ratios $E_{nc}/E_{nn}$ . The results from the scaling law given by 
Eq.~(\ref{s0}) are shown by the dot-dot-dashed line (lower one).
 {\it Right-frame:} Root mean square matter radius for $^{22}$C as a 
 function of $S_{2n}$ for the $n-c$ virtual state energy fixed at 1 MeV 
 (solid line) and 0 (dotted line). The $n-n$ virtual state energy is fixed to 143 keV. 
 The horizontal dashed lines represents the value of $\sqrt{\langle r^2_m\rangle}\,
 = \, 5.4 \pm 0.9$ fm extracted in the experiment of  Ref.~\cite{TanPRL10}.
}\label{fig:c22-2}
\end{figure}

\begin{figure}[h]
\vspace{1.2cm}
\begin{center}
\includegraphics[width=7.4 cm]{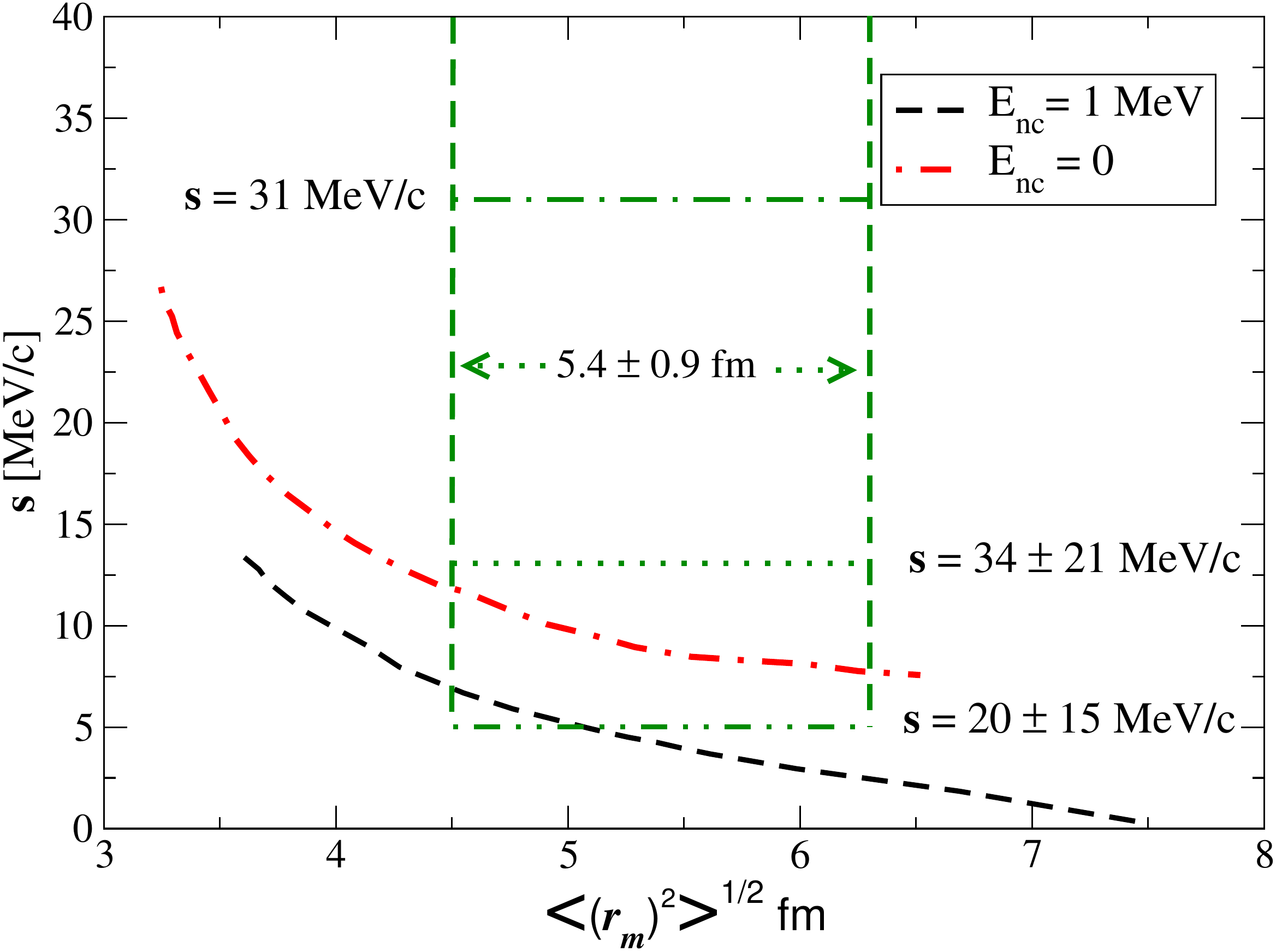}
\includegraphics[width=8. cm]{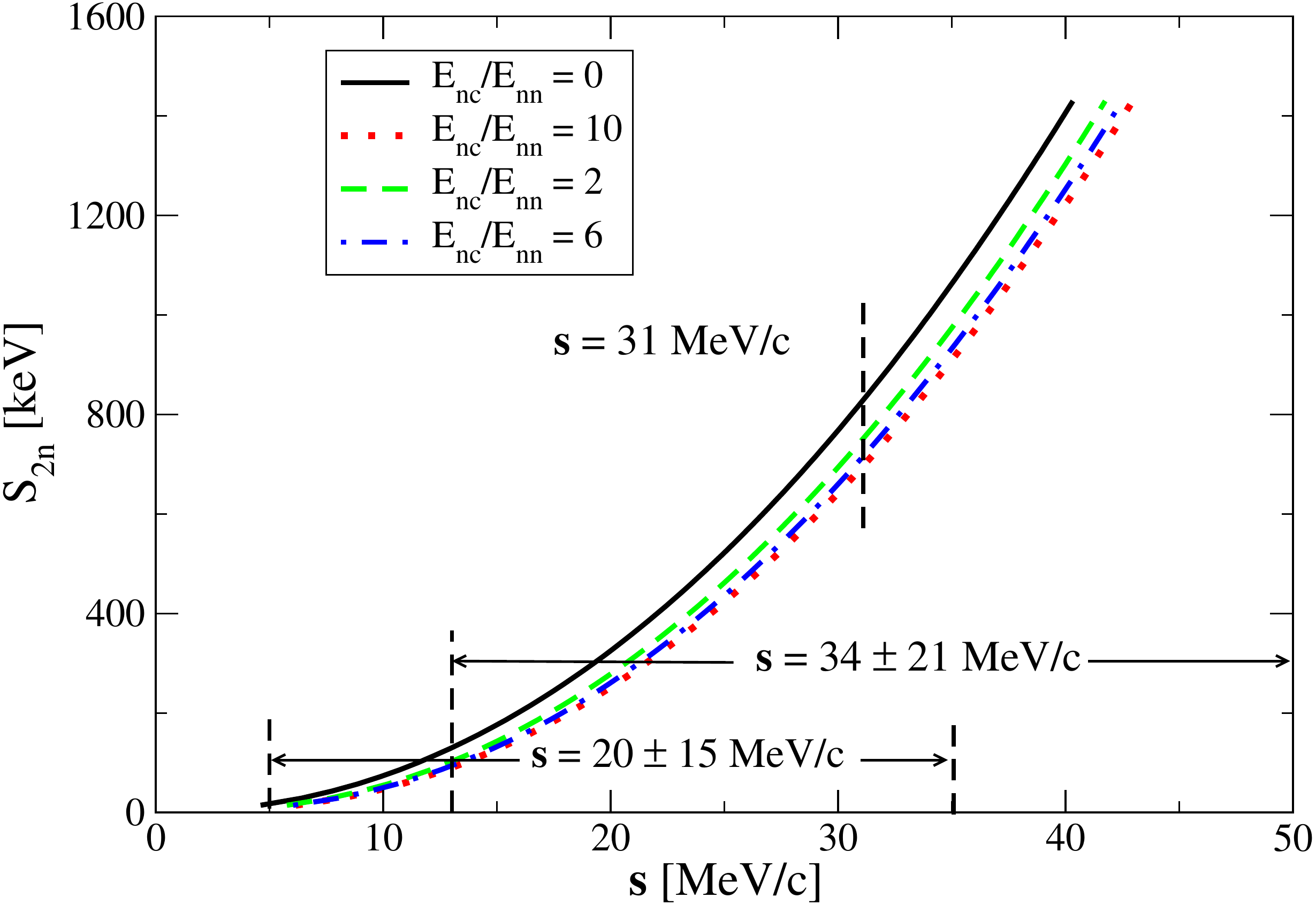}
\end{center}
\caption{(color online). {\it Left-frame:} Scaling  plot for the width ${\rm s}$ of the variance of
the core recoil momentum distribution in $^{22}$C  as
 a function function of the root mean square matter radius, 
 $\sqrt{\langle r_m^2\rangle}$, for  
 $E_{nc}=0$ (dot-dashed line) and for $E_{nc}=1$ MeV (dashed line), with a 
 fixed $n-n$ virtual state energy of 143 keV. 
 The extracted matter radius, $\sqrt{\langle r_m^2\rangle}\,= \, 5.4 \pm 0.9$ fm, from 
 Ref.~\cite{TanPRL10} is indicated in the figure.
 The values of ${\rm s}$ shown in the figure: 31 MeV/c \cite{KobPRC12} 
 (horizontal dot-dashed line), lower bound for ${\rm s}_{av}\,=\,20\,\pm\,15$ MeV/c
  (double-dot-dashed line ) derived from Eq. (\ref{d0}) and lower bound for
  ${\rm s}\,=\,34\,\pm\,21$ MeV/c (dotted line) obtained from Eq. (\ref{d1}).
 {\it Right-frame:} Two-neutron separation energy $S_{2n}$ in $^{22}$C as a function 
 of  ${\rm s}$ for fixed $n-n$ virtual state energy of 143 keV and different values of the 
 virtual $n-c$ energy. Calculations for
 $E_{nc}/E_{nn}$ values of 0 (solid line),  2 (dashed line), 6 (dot-dashed line)
  and  10 (dotted line).
}\label{fig:c22-3}
\end{figure}

\section{Scaling analysis $^{22}$C halo}\label{sec:analysis}

We apply the $n-n-c$ model to describe the halo of $^{22}$C and the 
results for the core recoil momentum distribution ${\rm s}$ expressed by the scaling 
function (\ref{sigmac}) are shown in the right-frame of Fig.~\ref{fig:c22-2}. We have kept fixed 
$E_{nn}=$ 143 keV and varied the energy of the 
virtual $s-$state of  $^{21}$C by considering  $E_{nc}/E_{nn}$ equal to 0, 2, 6 and 10. We 
allow a variation of $E_{nc}$, by including the region where it is 1 MeV, as indicated in 
Ref.~\cite{MosNPA13}, to spam different possibilities. The variation 
of $E_{nc}$ with fixed $S_{2n}$ and $E_{nn}$ does not show a dramatic effect, 
as we have pointed out in \cite{SouarXiv15}. We also present the scaling 
function at the unitary limit (\ref{s0}), and is quite close to 
the results with finite values of $E_{nc}$ and $E_{nn}$, showing very clearly 
in the model that $S_{2n}$ is the dominating scale in the core recoil 
momentum distribution in the three-body halo at low momentum. In addition,  
for large values of $S_{2n}$, i.e.,  when $S_{2n}$ is much larger than $E_{nn}$ 
and $E_{nc}$, the scaling function tends to approach the unitary result, as expected.

The matter radius is another quantity that we can compute from the halo neutron 
orbit radius with respect to the center of mass, $\sqrt{\langle r^2_n \rangle}$, which
can also be written in terms of a scaling function~\cite{YamNPA04}, such that
\begin{equation}\label{rn}
\sqrt{\langle r^2_n \rangle}=\sqrt{\frac{m_n}{\hbar^2}\,S_{2n}}\, \mathcal{ R}_n\left(\pm
\sqrt{\frac{E_{nn}}{S_{2n}}},\,
\pm\sqrt{\frac{E_{nc}}{S_{2n}}}; {A}\right)\, .
\end{equation}
This allows us to compute the matter radius, as
\begin{equation}\label{rm}
\sqrt{\langle r^2_m[^{22}C] \rangle}=\sqrt{\frac{2}{22}\langle r_n^2\rangle+
\frac{20}{22}\langle r_m^2[^{20}C]\rangle}.
\end{equation}
Only by considering the $^{22}$C extracted matter radius of 5.4(9) fm \cite{TanPRL10},
the conclusion of \cite{YamPLB11}, based on the model we are following here, was that 
$S_{2n}$ is below 100 keV. However, in view of the new data on the core momentum distribution, 
we ought to include it on the analysis of the two-neutron separation energy. 
The right frame of Fig.~\ref{fig:c22-2} summarize the findings of \cite{YamPLB11}, where 
we plot the correlation between $\sqrt{\langle r_m^2\rangle}$ and $S_{2n}$. 
The virtual $s-$ state energy of $^{21}$C is chosen to be 0 and 1 MeV \cite{MosNPA13}, with 
fixed $E_{nn}=143$. In the case of $E_{nc}=0$, 
the matter radius is compatible with $S_{2n}\,<\, 100$ keV, while by increasing the
$E_{nc}$ the upper limit for $S_{2n}$ decreases, as shown by the results obtained for $E_{nc}=1$ 
MeV. This happens because the interaction becomes weaker as the $E_{nc}$ increases;
and, for a given three-body binding energy, the system has to shrink \cite{YamNPA04}. This is the 
effect behind the decrease of the possible values of $S_{2n}$, when the $^{21}$C virtual state energy 
is varied from 0 to 1 MeV.

We have data extracted from experiments for the width ${\rm s}$ of the core recoil momentum 
distribution and the matter radius of $^{22}$C. Then, we just built the correlation 
 between  these quantities and analyse the data in view of the scaling  function of our 
 model. The left-frame of Fig.~\ref{fig:c22-3} shows the correlation between the rms matter
 radius and  ${\rm s}$ for the values of $E_{nc}$ equal to 0 and 1 MeV. For a given matter 
radius ${\rm s}$ increase by changing $E_{nc}$ from 0  of to 1 MeV. The 
variation is noticeable, about a factor of 2. That effect is a reminiscent 
of the contraction of the wave function by decreasing the energy $E_{nc}$. The region
delimited by the matter radius of $5.4\pm.9$fm and the three 
values for ${\rm s}$, namely, 31 MeV/c \cite{KobPRC12}, 20$\pm$15 MeV/c from 
the fit (\ref{d0}) and 34$\pm$21 MeV/c from the fit (\ref{d1}) is shown in the 
left-frame of the figure. It looks that the value of $E_{nc}=$ 1 MeV is too large from
the consistency between the scaling function, the matter radius and the lower bound of 
${\rm s} =$ 20$\pm$15 MeV/c. Just by adopting ${\rm s}=$ 31 MeV/c the matter radius basically
goes to the core radius $\sim$ 3 fm. 
The lower bounds of 20$\pm$15 MeV/c 
and 34$\pm$21 MeV/c, taken together with the scaling function seems consistent 
with the matter radius. However, the central values of $\rm s$ suggests that the matter radius is overestimated. In order to be consistent with the matter radius the distribution 
should be quite narrow according to the scaling function.

The right-frame of Fig.~\ref{fig:c22-3} completes our analysis of the two-neutron
separation energy in $^{22}$C. We show $S_{2n}$ as a function of ${\rm s}$. The data 
values for ${\rm s}$ suggest that $S_{2n}$ could have larger values than estimated before, 
and stay above 100 keV. Taking as reference the data with the average between 
positive and negative momentum around the maximum  of core 
momentum distribution peak, i.e., ${\rm s}=$ 20$\pm$15 MeV/c, one 
gets $S_{2n}\sim$ 400 keV and $\sqrt{\langle r_m^2\rangle} \sim$ 3.5 fm. The values of ${\rm s}$  equal to 
31 MeV/c \cite{KobPRC12} and 34$\pm$21 MeV/c (\ref{d1}) suggest much larger values 
of the two-neutron separation energy about 800 keV and 900 keV, respectively.
The neutron halo of $^{22}$C in this case would be quite small compared to the
corresponding one in $^{11}$Li, and possibly would not justify the increase of the
 reaction cross section of $^{22}$C compared to the neighbour 
nuclei, $^{19}$C and $^{20}$C, as observed by Tanaka and collaborators 
\cite{TanPRL10}, in the collision of $^{22}$C on a liquid hydrogen target. 
This fact together with the clear presence of the narrow peak in 
the core recoil momentum distribution point our analysis to the value of 
${\rm s} < $ 20 Mev/c, $S_{2n}<$ 400 keV and $\sqrt{\langle r_m^2\rangle} >$ 3.5 fm. 
The sensitivity to the virtual $s-$state energy of $^{21}$C is quite small in ${\rm s}$ 
for a given $S_{2n}$, but it has a noticeable effect in the matter 
radius of $^{22}$C shrinking considerably the halo as $E_{nc}$ runs from 0 to 1 MeV. 
That tension between the halo size and the virtual state of $^{21}$C indicates that 
smaller values of $E_{nc}$ below 1 MeV are preferred by our analysis. We 
rely on the dominance of the Efimov physics on the halo properties and in the
Serber model to interpret the data on the core recoil momentum distribution.

\section{Conclusions}

The core recoil momentum distribution of Borromean 
neutron-rich isotopes of light exotic nuclei is studied within a three-body 
model constituted by a core and two neutrons, where the interaction is dominated by 
the $s-$wave channel. We adopt a renormalized  three-body model with a zero-range 
force, that naturally accounts for the universal Efimov physics. Our model
is adequate to describe a two-neutron halo weakly bound to a 
core nuclei, where these neutrons explore with a high probability the 
classically forbidden region well outside the core. In addition, 
the $n-c$ scattering length should be larger than the interaction range.  
These conditions are well satisfied by the Borromean halo of $^{11}$Li, the 
prototype of a weakly-bound radioactive halo nuclei at the neutron drip line.

The core momentum distribution was calculated from the halo wave function in 
momentum space, obtained from the numerical 
solution of the subtracted homogeneous Skorniakov and Ter-Martorisyan
coupled equations for the zero-range interaction in the the mass imbalanced 
case. The inputs are the two-neutron separation energy and the energies 
of the singlet $n-n$  and $n-c$ virtual states. 
We calculated the momentum probability densities for the
Borromean exotic nuclei $^{11}$Li, $^{14}$Be and $^{22}$C. A fair reproduction 
of the experimental data was obtained in the case of  the core recoil 
momentum distribution of  $^{11}$Li and $^{14}$Be without resorting to 
free parameters.

We suggest  a simple method to separate out the cross-section data for the
fragmentation of the $^{22}$C the narrow peak in the 
core recoil momentum distribution associated with the breakup of the 
halo structure. For that aim, we performed the subtraction of a wide distribution 
associated with the core recoil momentum momentum distribution observed for 
$^{20}$C. Indeed, the data for the fragmentation differential cross-section for the core 
momentum above 100 MeV/c in $^{22}$C and $^{20}$C seems indistinguishable, which 
indicates that the two neutrons in the wide part of the distribution come 
from inner part of the orbits localized within the core. The inverse of the 
difference between those cross-sections, after averaging them for 
positive and negative momentum, present a linear behaviour 
with $p^2$, which allows to extract the width of the momentum 
distribution. 
We apply the same fitting procedure to the difference of cross-sections but without the 
average. With these values at hand, we proceed to make a combined analysis of 
the width of the core recoil momentum distribution and matter radius for $^{22}$C. 
 
We computed the scaling functions correlating  ${\rm s}$, $S_{2n}$ 
and the virtual state energies of the two-body subsystems, and also the scaling functions 
correlating  the matter radius with these quantities. 
From the scaling function correlating ${\rm s}$ and $S_{2n}$ and 
the experimental data for ${\rm s}$ equal to 31 MeV/c \cite{KobPRC12}, 
20$\pm$15 MeV/c and 34$\pm$21 MeV/c, it is suggested that $S_{2n}$ 
could be well above 100 keV larger than previous estimations. 
From the correlation between the matter radius and ${\rm s}$, and the
reference the value  of ${\rm s}_{av}\,=\, 20\pm15$ MeV/c, one gets: 
$\sqrt{\langle r_m^2\rangle}\sim$ 3.5 fm and $S_{2n}\sim$ 400 keV. 
However, the values of 31 MeV/c and 34$\pm$21 MeV/c (\ref{d1}) suggest much larger values 
of the two-neutron separation energy, about 800 keV and 900 keV, respectively.
In these last two cases, the halo of $^{22}$C would be quite small compared to the
corresponding one in $^{11}$Li, and we believe that it would not justify the increase of 
the reaction cross section of $^{22}$C compared to the neighbour nuclei, 
$^{19}$C and $^{20}$C, as the experiment by Tanaka and collaborators~\cite{TanPRL10} is showing. In addition, by considering that it is  
clear the evidence of the presence of a narrow peak in the core recoil momentum distribution, as our 
analysis, model independent, is pointing out, it seems preferential that 
${\rm s} < $ 20 Mev/c, $S_{2n}<$ 400 keV and $\sqrt{\langle r_m^2\rangle}>$ 3.5 fm.

Our study also shows that the sensitivity of the width of the $^{22}$C core momentum distribution
to the virtual $s-$state energy of $^{21}$C is quite small for $S_{2n}$ fixed, but the matter radius is quite 
sensitive to it. The variation of the virtual state of  $^{21}$C from 0 to 1 MeV shrinks considerably the 
matter radius.  Such a tension between the halo size and the virtual state of $^{21}$C suggests, within
our analysis, that smaller values of $E_{nc}$ below 1 MeV are preferred. Such conclusion is based on 
the belief that the Efimov physics, expressed by the calculated scaling laws, dominates the halo 
properties, and also by considering the validity of the Serber model to interpret the data on the core 
recoil momentum distribution.
In short, the combined analysis of the core momentum distribution and matter 
radius for $^{22}$C, on the basis of the scaling laws computed close to the Efimov 
limit, suggest: 
(i) a virtual state of $^{21}$C well below 1 MeV; 
(ii) the overestimation of the extracted matter $^{22}$C radius; and 
(iii) a two-neutron separation energy of this halo nucleus to a value between 100 and 400 keV.

\section*{Acknowledgement}
This work was partially supported by the Brazilian agencies CAPES, FAPESP and CNPq.

\end{document}